\documentclass[aps,prl,twocolumn,showpacs,superscriptaddress]{revtex4-1}
\usepackage{graphicx,amsmath,amssymb,amsfonts}
\usepackage[latin1]{inputenc}
\usepackage{array}
\usepackage{multirow}
\usepackage{float}
\usepackage{color}
\usepackage[normalem]{ulem}
\begin{document}

\title{DyOCl: a rare-earth based two-dimensional van der Waals material with strong magnetic anisotropy}

\affiliation{Beijing Key Laboratory of Optoelectronic Functional Materials and MicroNano
	Devices, Department of Physics, Renmin University of China,
	Beijing 100872, China}
\affiliation{Quzhou University, Quzhou, Zhejiang 32400, China}

\author{Congkuan Tian}
\thanks{These authors contributed equally to this work}
\affiliation{Beijing Key Laboratory of Optoelectronic Functional Materials and MicroNano
	Devices, Department of Physics, Renmin University of China,
	Beijing 100872, China}

\affiliation{International Center for Quantum Materials, School of Physics, Peking University, Beijing 100871, China}
\affiliation{Beijing Academy of Quantum Information Sciences, Beijing 100193, China}

\author{Feihao Pan}
\thanks{These authors contributed equally to this work}
\author{Le Wang}
\affiliation{Beijing Key Laboratory of Optoelectronic Functional Materials and MicroNano
	Devices, Department of Physics, Renmin University of China,
	Beijing 100872, China}

\author{Dehua Ye}
\affiliation{Quzhou University, Quzhou, Zhejiang 32400, China}

\author{Jieming Sheng}
\affiliation{ Institute of High Energy Physics, Chinese
	Academy of
	Sciences (CAS), Beijing 100049, China }
\affiliation{ Spallation
	Neutron Source Science Center, Dongguan 523803, China}
\affiliation{ Department of Physics, Southern University of Science
	and Technology, Shenzhen 518055, China}

\author{Jinchen Wang}
\author{Juanjuan Liu}
\author{Jiale Huang}
\affiliation{Beijing Key Laboratory of Optoelectronic Functional Materials and MicroNano
    Devices, Department of Physics, Renmin University of China,
    Beijing 100872, China}

\author{Hongxia Zhang}
\author{Daye Xu}
\affiliation{Beijing Key Laboratory of Optoelectronic Functional Materials and MicroNano
	Devices, Department of Physics, Renmin University of China,
	Beijing 100872, China}

\author{Jianfei Qin}
\author{Lijie Hao}
\affiliation{ China Institute of Atomic Energy, PO Box-275-30, Beijing 102413, China}

\author{Yuanhua Xia}
\author{Hao Li}
\affiliation{ Key Laboratory of Neutron Physics and Institute of
	Nuclear Physics and Chemistry, China Academy of Engineering
	Physics, Mianyang 621999, China}
\author{Xin Tong}
\affiliation{ Institute of High Energy Physics, Chinese
	Academy of
	Sciences (CAS), Beijing 100049, China }
\affiliation{ Spallation
	Neutron Source Science Center, Dongguan 523803, China}

\author{Liusuo Wu}
\affiliation{ Department of Physics, Southern University of Science
	and Technology, Shenzhen 518055, China}

\author{Jian-Hao Chen}
\author{Shuang Jia}
\affiliation{International Center for Quantum Materials, School of Physics, Peking University, Beijing 100871, China}
\affiliation{Beijing Academy of Quantum Information Sciences, Beijing 100193, China}

\author{Peng Cheng}
\email[Corresponding author: ]{pcheng@ruc.edu.cn}
\affiliation{Beijing Key Laboratory of Optoelectronic Functional Materials and MicroNano
	Devices, Department of Physics, Renmin University of China,
	Beijing 100872, China}

\author{Jianhui Yang}
\email[Corresponding author: ]{yangjh@qzc.edu.cn}
\author{Youqu Zheng}
\email[Corresponding author: ]{zyq@qzc.edu.cn}
\affiliation{Quzhou University, Quzhou, Zhejiang 32400, China}

\begin{abstract}
Comparing with the widely known transitional metal based van der Waals (vdW) materials, rare-earth based ones are rarely explored in the research of intrinsic two-dimensional (2D) magnetism. In this work, we report the physical properties of DyOCl, a rare-earth based vdW magnetic insulator with direct band gap of $\sim 5.72~eV$. The magnetic order of bulk DyOCl is determined by neutron scattering as the $A$-type antiferromagnetic structure below the N\'{e}el temperature $T_N=10~$K. The large magnetic moment near 10.1~$ \mu_{B} $/Dy lies parallel to the $a$-axis with strong uniaxial magnetic anisotropy. At $2~K$, a moderate magnetic field ($\sim 2~T$) applied along the easy axis generates spin-flip transitions and polarizes DyOCl to a ferromagnetic state. Density functional theory calculations reveal an extremely large magnetic anisotropy energy ($-5850~\mu eV/Dy$) for DyOCl, indicating the great potentials to realize magnetism in 2D limit. Furthermore, the mechanical exfoliation of bulk DyOCl single crystals down to seven layers is demonstrated. Our findings suggest DyOCl is a promising material playground to investigate 2D $f$-electron magnetism and spintronic applications at the nanoscale.
\end{abstract}

\maketitle




\section{Introduction}
As the parent compounds of atomically thin magnet, two-dimensional (2D) van der Waals (vdW) magnetic materials have received a huge attention since the discoveries of 2D intrinsic magnetism in CrI$_{3}$\cite{CrI3} and Gr$_{2}$Ge$_{2}$Te$_{6}$\cite{Cr2Ge2Te6}. These cleavable materials, including both ferromagnet and antiferromagnet, can sustain magnetic order down to mono- or few-layer thickness. They are important for both spintronic applications and exploring exotic quantum phases in condensed matter physics\cite{Burch2018}. The intensive researches in recent years have lead to the discoveries of tunneling magnetoresistance\cite{MR}, large anomalous Hall effect\cite{AHE}, magnetic skyrmions\cite{Sky1}, heavy fermion state\cite{2018Kondo} and topological spin excitations\cite{CrI3Neutron} in this kind of materials.

\begin{figure*}[htbp]
	\centering
	\includegraphics[width=\textwidth]{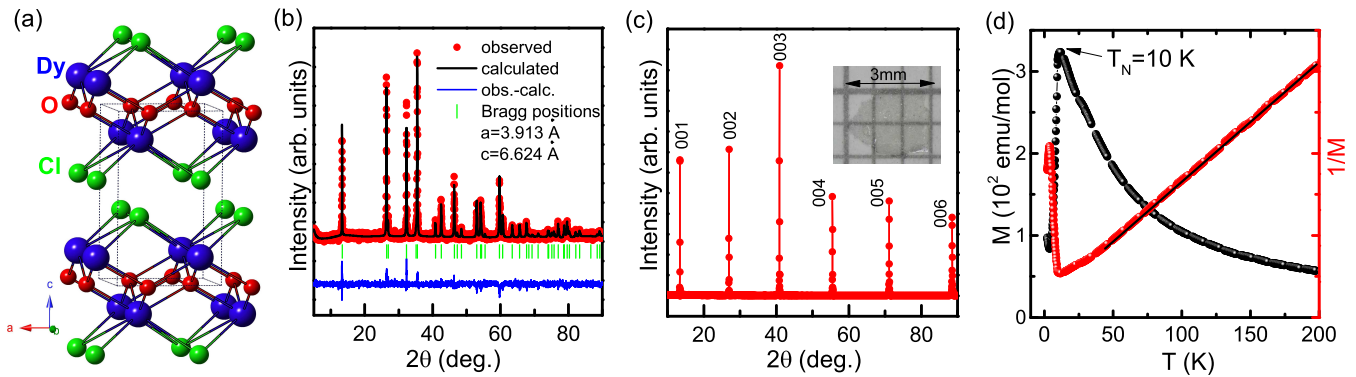}
	\caption {(a): Crystal structure of DyOCl. The vdW gap exists between adjacent Cl-layers. (b) Powder X-ray diffraction data for DyOCl with refined lattice parameters provided. (c) X-ray diffraction patterns from the $ab$-plane of a single crystal. The inset shows a picture of one transparent DyOCl single crystal. (d) Temperature-dependent magnetization of polycrystalline samples of DyOCl measured with $H=1~ kOe$. The black solid line shows the result of Curie-Weiss fit. } \label{fig1}
\end{figure*}

So far, the investigations on 2D magnetism are limited to a few examples, such as FePS$_{3}$\cite{FePS3}, CrX$_{3}$ (X=I, Br, Cl)\cite{CrI3,CrX3}, VI$_{3}$\cite{VI3}, Cr$_{2}$Ge$_{2}$Te$_{6}$\cite{Cr2Ge2Te6} and Fe$_{n}$GeTe$_{2}$ (n=3,4,5)\cite{Fe3GeTe2_1,Fe3GeTe2_2,Fe4GeTe2,Fe51,Fe52}. Almost all of them are transitional metal compounds while rare-earth (RE) based 2D materials have been rarely reported. Although the stronger exchange interactions between transitional metal ions are favored in 2D spintronic applications, RE based materials may also have special advantages in such research field. 

Theoretically, in order to make spin fluctuations finite and stabilize long-range magnetic order in 2D limit, it is essential for a gap opening in the spin-wave spectrum\cite{SWT1}. Strong magnetic anisotropy and dipolar interaction are the two main factors for introducing such a gap\cite{SWT1,SWT2}. From this point of view, rare-earth vdW materials may have advantages in realizing 2D magnetism since many RE ions tend to exhibit strong Ising magnetic anisotropy due to the spin-orbit coupling and crystal field effect\cite{Tb2Ti2O7,DyScO3}. On the other hand, comparing with transitional metal ions, many RE ions have much larger magnetic moments and consequently bring stronger magnetic dipolar interactions. Furthermore the $4f$-electron magnetism of RE vdW materials might introduce new phenomena and physics different from the intensively investigated $3d$-electron magnetism in 2D materials.

Although the studies on RE-based vdW materials are very few, some intriguing discoveries have been reported. The initial investigation on EuSi$_{2}$ and GdSi$_{2}$ has revealed the evolution from the bulk antiferromagnetism to intrinsic 2D ferromagnetism of ultra-thin layers\cite{EuSi2}. A recent study found that GdTe$_{3}$ is the first vdW material with both high mobility and magnetism, which might provide potential for new twistronic and spintronic elecntrical applications\cite{GdTe3}. Very recently, YbCl$_3$\cite{YbCl3} and YbOCl\cite{ZhangQM} were proposed to be Kitaev spin liquid candidates. Besides these reports, the investigation on magnetic RE vdW materials is still at the early stage.
  
In this paper, we introduce a RE-based magnetic vdW material DyOCl, which is an antiferromagnet with $T_N=10~$K. The $A$-type antiferromagnetic (AFM) structure, strong uniaxial magnetic anisotropy, large magnetic moment and field-induced spin-flip transitions of DyOCl are determined experimentally, consistent with the large magnetic anisotropy energy calculated from density functional theory. These intriguing magnetic properties, combined with the availability of exfoliating DyOCl single crystals down to few layers, suggest DyOCl is a promising candidate for further investigations on $4f$-electron magnetism in 2D limit.

\section{methods}

Polycrystalline samples of DyOCl were synthesized by
solid-state reaction of NH$_{4}$Cl and Dy$_{2}$O$_{3}$ powder in the mole ratio 3 : 1. These reagents were mixed and heated to $650\,^{\circ}\mathrm{C}$ for $1.5~h$ in an alumina crucible, then cooled with the furnace. Single crystals were grown by flux method with Dy$_{2}$O$_{3}$: DyCl$_{3}$ (1 : 15) in an alumina crucible enclosed in a evacuated quartz tube. The mixture was heated to $1100\,^{\circ}\mathrm{C}$ and maintained at this temperature for $24~h$ before it was slow-cooled to $700\,^{\circ}\mathrm{C}$ at a rate of $2\,^{\circ}\mathrm{C}/h$. 
Transparent single crystals with dimension up to $3 \times 2 \times 0.05~mm^{3}$ [inset of Fig. 1(c)] could be obtained and excess flux can be dissolved by water. All the samples are air-stable.

X-ray diffraction (XRD) patterns of powder samples were collected
from a Bruker D8 Advance X-ray diffractometer using Cu
K$_{\alpha}$ radiation. Magnetization measurements were carried
out in Quantum Design MPMS3 and PPMS-14T. The powder neutron diffraction experiments were carried out on Xingzhi cold neutron triple-axis spectrometer at the China Advanced Research Reactor (CARR) and Xuanwu powder neutron diffraction spectrometer at China Academy of Engineering Physics (CAEP). For neutron experiments on Xingzhi (Data presented in Fig.2), the incident neutron energy was fixed at $15$~meV with a neutron velocity selector used upstream to remove higher order neutrons\cite{XingZhi}. Approximate $4~$g of DyOCl powder sample sealed in a cylindrical aluminium container was loaded into a closed cycle refrigerator that regulates the temperature from $3.5~$K to $300~$K. Because the dysprosium (Dy) is highly neutron absorbing, the absorption correction was applied to neutron powder diffraction data. The powder sample was regulated into a cylinder-shape with diameter of 10~mm and absorption corrections were calculated based on this shape. The program FullProf Suite package was used in the Rietveld refinement of neutron powder diffraction data\cite{RODRIGUEZCARVAJAL199355,Rietveld.a07067}. No preferred crystal orientation due to the texture effects is revealed from the refinement. The diffusion reflectance spectroscopy was measured on a Shimadzu UV-3600 UV-VIS-NIR spectrophotometer. The dimensions of exfoliated DyOCl nanoflakes were checked by a Bruker edge dimension atomic force microscope. 

\begin{figure*}[htbp]
	\centering
	\includegraphics[width=15cm]{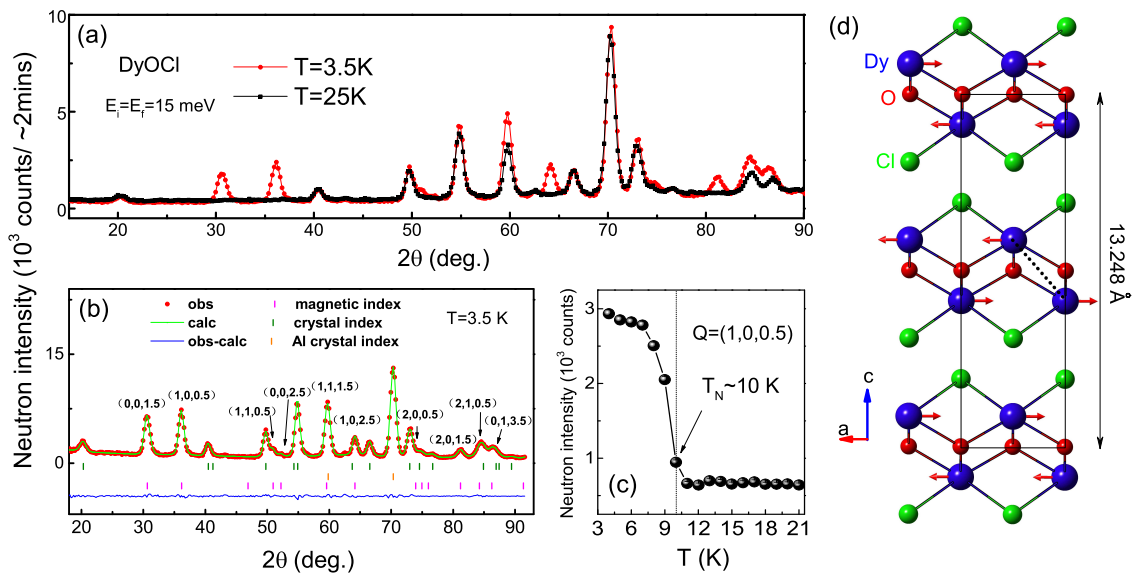}
	\caption {(a) Powder neutron diffraction patterns of DyOCl at $T=3.5~K$ and $T=25~K$. (b) The best fit of the diffraction data at $3.5~K$ is achieved by the magnetic structure depicted in (d). The aluminum peaks come from the sample can. (c) The temperature dependent intensities of magnetic peak (1,0,0.5). (d) Antiferromagnetic structure of DyOCl at $3.5~K$. The magnetic unit cell is marked by the black rectangle. Since the moments of Dy are aligned ferromagnetically in the $ab$-plane, only the spin configuration viewed on $ac$-plane is presented.} \label{fig2}
\end{figure*}

Spin polarized density functional theory (DFT) was used for all calculations including ionic relaxation, which is integrated in the Vienna ab initio simulation package (VASP)\cite{T1,T2,T3}. The generalized gradient approximation (GGA) with the Perdew-Burke-Ernzerhof (PBE) functional was used to determine the exchange-correlation energy\cite{T4}. The kinetic-energy cutoff was set to $500~eV$. Projector augmented wave (PAW) potentials were used for all calculations\cite{T5}. Ionic relaxation was performed until the force on each atom was below $0.01~eV/\AA$. The density of K-points in real space was less than $0.03\times0.03\times0.03\AA^{-3}$ for all calculations, based on the Monkhorst-Pack method\cite{T6}. Non-Collinear GGA+U calculation was implemented with the parameters $U_{eff}$ as $6.0~eV$ for Dy element\cite{T7}. For band structure calculations, the $4f$-electrons of Dy element were regard as core electron. The binding energy between different layers was calculated by using the dispersion correction (DFT-D3) method\cite{T8}.

\section{Results and discussions}

\subsection{Crystal and magnetic structure}

The rare-earth oxychloride DyOCl have PbFCl-type structure which belongs to the tetragonal crystal system with $P4/nmm$ (No.129) as the space group [Fig. 1(a)]\cite{DyOCl2002}. Its crystal structure is different from YbOCl, which has a SmSI-type structure with honeycomb lattice\cite{ZhangQM}. This crystal structure is confirmed by X-ray analysis of both powders diffraction patterns and $00l$ relections of single crystal, as presented in Fig. 1(b) and (c). All samples were found to be phase pure within the instrumental resolutions. The lattice parameters obtained from Rietveld refinement are $a=b=3.913$ \text{\AA} and $c=6.624$ \text{\AA}. According to our DFT calculations, the vdW gap exists between Cl-layers with cleavage energy of $36.8~meV/A^2$, comparable with that of Fe$_{3}$GeTe$_{2}$. Therefore, few-layer DyOCl is expected to be cleaved from its bulk counterpart.

To the best of our knowledge, although the luminescent and catalytic properties of rare-earth oxychlorides have been widely studied, their magnetic properties were rarely investigated. The only knowledge about the magnetism of DyOCl previously is that it undergoes an AFM-like transition at around $T=11~K$\cite{DyOCl}. Our magnetic susceptibility measurements of DyOCl powder confirm this AFM transition at $T_{N}=10~K$ as shown in Fig. 1(d). The Curie-Weiss (CW) fit of the high-temperature susceptibility data ($T\textgreater50~K$) yields effective moment $\mu_{eff}/Dy=10.3~\mu_{B}$ and CW temperature $\theta_{CW}=-22.9~K$. It should be noted that the fitting result of the CW temperature is likely impacted by crystal field effect. Its physical meaning need to be clarified based on future determination about the crystal field levels of DyOCl.  

\begin{figure}
	\includegraphics[width=6.7cm]{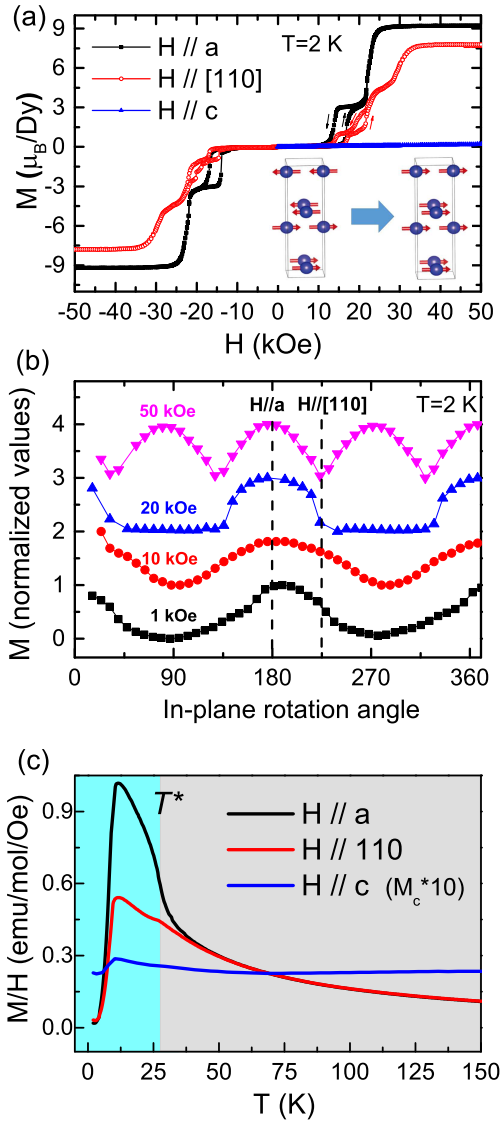}
	\caption {(a) The field-dependent magnetization measured at $2~K$ of a DyOCl single crystal with applied field along the $a$, $[110]$ and $c$ axes. The inset shows the spin configurations of Dy$^{3+}$ in one unit cell transformed from AFM state at low field to FM state at high field. (b) $ab$-plane angle-dependent magnetization measured at $2~K$ and magnetic field from $1~kOe$ to $50~kOe$. (c) The temperature-dependent magnetization along three different crystalline axes. The magnetization along $c$-axis is multiplied by a factor of 10. For in-plane magnetization, a weak kink appears at $T^*$ with significant magnetic anisotropy emerging below this temperature. } \label{fig3}
\end{figure}

Powder neutron diffraction on DyOCl was performed to determine the low temperature AFM structure. Fig. 2 (a) shows the raw data collected at $T=3.5~K$ and $T=25~K$. Comparing with the data at $25~K$, the appearance of new peaks and the large intensity enhancement of certain peaks at $3.5~K$ suggest the notable contributions from magnetic scattering. These magnetic Bragg peaks can be well indexed as marked in Fig. 2(b). The occurrence of an AFM transition at $T_{N}=10~K$ is also demonstrated by the temperature dependent intensities of magnetic Bragg peak $(1,0,0.5)$ [Fig. 2(c)]. All the magnetic Bragg peaks indexed could be well defined by the propagation vector $k=(0,0,0.5)$,  which means the magnetic unit cell is doubled along $c$-axis in respect to the crystal unit cell. Through the representation analysis using FullProf, only four types of magnetic structures are compatible with this propagation vector. After absorption correction, the diffraction data at $T=3.5~K$ are best fitted by the magnetic structure depicted in Fig. 2(d), the fitting results are shown in Fig. 2(b) with R$ _{p}$=3.35, R$ _{wp}$=4.41 and $ \chi^{2} $=4.53. The fit of other three types of magnetic structures yields unacceptable high R$ _{p}$ and R$ _{wp}$ factors. 

Therefore, DyOCl develops an $A$-type AFM structure below $T_{N}=10~K$. The moments of Dy are confined within the easy $ab$-plane and ferromagnetically aligned in a single plane of Dy. For one vdW layer of DyOCl, it contains two Dy planes which are antiferromagnetically coupled and separated by oxygen atoms. The nearest Dy-Dy distance is 3.587\text{\AA}, which is indicated by the dotted line in Fig. 2(d). This magnetic structure suggest a nearest-neighbor AFM interaction of Dy and the potential monolayer DyOCl still has an AFM ground state. The ordered moment obtained from Rietveld refinement is $ \mu $=10.1 $ \mu_{B} $/Dy$ ^{3+} $ at $3.5~ K$, which is almost the same as the saturation moment of a free Dy$^{3+}$ ($\sim10\mu_{B}$). Although the in-plane moment direction cannot be resolved from powder diffraction data due to tetragonal structure of DyOCl, it is confirmed to be along $a$-axis through the anisotropic magnetization measurements on single crystals as demonstrated in the next section.

\subsection{Anisotropic magnetic properties}

Since DyOCl adopts a tetragonal lattice symmetry and the point symmetry of the Dy site is $C_{4v}$, the isothermal magnetization measurements were carried out along three principal crystallographic axes, namely $a$, $[110]$ and $c$ axes respectively. The results at $2~K$ are shown in Fig. 3(a). For field-dependent magnetization along $a$-axis, the initial magnetization shows a very weak increase versus field which is consistent with an AFM state. At $H\textgreater 15~kOe$, two successive magnetization-jumps appear and the magnetization becomes saturated for $H\textgreater 25~kOe$. The saturated moment per Dy is $9.2~\mu_{B}$, which is quite close to the theoretical value $\mu_J \sqrt{J/(J+1)}$ in a localized model. Therefore DyOCl can be fully polarized to a ferromagnetic state from the AFM ground state with a moderate magnetic field. This spin-flip transition is illustrated in the inset of Fig.3 (a) and have additional features as revealed by the hysteresis loops. For $H\parallel a$, the magnetization $M_{a}(H)$ undergoes two steep magnetization jumps at $H\approx16~kOe$ and $H\approx22~kOe$. The first jump exhibits a large hysteresis, indicating a first-order transition. In contrast, no hysteresis appears for the second jump. There is a magnetization plateau after the first jump. The moment at this plateau is $3.1~\mu_{B}/Dy$ which is one third of the saturation moment. These features could be associated with possible spin-flop transitions, magnetic quantum tunneling between different magnetic states of Dy ion or other meta-magnetic transitions.

On the other hand, significant magnetic anisotropy is observed between the $ab$-plane and $c$-axis data. The magnetization along $c$-axis at $50~kOe$ is only $0.2~\mu_{B}$, which is 46 times smaller than that along the $a$-axis. Strikingly, DyOCl also exhibits strong magnetic anisotropy within $ab$-plane. The magnetization along $[110]$-axis has much larger saturation field ($3.5~kOe$) and smaller saturated moment ($7.7~\mu_{B}$), in contrast to that along $a$-axis. In addition, the $M_{110}(H)$ curve has more complex field-induced jumps compared with that in $M_{a}(H)$.

The magnetic anisotropy within $ab$-plane can be further illustrated by angle-dependent magnetization measurements. As shown in Fig. 3(b), by varying direction of the field relative to the sample in the $ab$-plane, the maximum magnetization appears at $H\parallel a$. Therefore the magnetic easy axis of Dy is the $a$-axis. Since the tetragonal lattice and crystalline electrical field of DyOCl have four-fold symmetry, the $a$-axis should be equivalent to $b$-axis. However under the small applied field, up to $20~kOe$, the in-plane magnetization has a two-fold symmetry with maximum along $a$- and minimum along $b$-axis. There are two reasons for such a result. Firstly, the initial field was applied near $a$-axis during the measurement, thus DyOCl ordered with moments lying along $a$-axis. Secondly, magnetic field below $20~kOe$ is not enough to tune the moment direction freely between equivalent $a$- and $b$-axis. When higher magnetic field of $50~kOe$ was applied, a four-fold symmetric pattern arises in the in-plane magnetization with maximum along both $a$- and $b$-axis, minimum along $[110]$-axis. The above results demonstrate the large in-plane magnetic anisotropy of DyOCl, which is quite different from 2D transitional metal based magnetic vdW magnets with almost negligible in-plane anisotropy such as CrCl$_{3}$ and Co-doped Fe-Ge-Te series\cite{CrCl3,PCheng_APL2020,Fe5May,Fe4Co}. Therefore, although DyOCl favors an in-plane magnetization, it actually exhibits an Ising-like uniaxial magnetocrystalline anisotropy which enables the opening a spin-wave gap to resist thermal fluctuations.

Fig. 3(c) shows the temperature dependent magnetization along different crystal directions. Besides the AFM-like cusp at $T_{N}$ in all $M(T)$ curves, a striking feature is that significant in-plane magnetic anisotropy actually appears at much higher temperature than $T_{N}$. $M_{a}$ and $M_{[110]}$ almost overlap above $T^{*}=27~K$, but they are well separated below $T^{*}$. At $12~K$, $M_{a}$ is almost twice as large as $M_{110}$. In addition, there are also small kinks at around $T^{*}$ for both $M_{a}$ and $M_{110}$. This indicates possible new phase transitions or the development of short-range magnetic orders at $T^{*}$. However our current neutron scattering measurements did not find any clues for additional transitions, further high-resolution neutron or X-ray investigations are needed for clarifications.   

Magnetic anisotropy energy ($E_{MAE}$) of DyOCl is calculated through density functional theory. $E_{MAE}$ is defined as the energy difference between the easy- and hard-axis magnetizations, which is a critical parameter to stabilize the long-range magnetic order against thermal fluctuations in 2D materials\cite{T9,T10}. In principal, $E_{MAE}$ contains the magnetocrystalline anisotropy energy ($E_{MCA}$) induced by the spin-orbit coupling and magnetic shape anisotropy energy ($E_{MSA}$) due to the magnetic dipole-dipole interaction\cite{T9,T10}. For DyOCl, our calculations reveal that $E_{MSA}$ ($ -4.8~\mu eV/Dy$) is negligible small. However the values of $E_{MCA}$ is $ -5850~\mu eV/Dy$, which is a surprisingly large value comparing with that of transitional metal based 2D vdW magnetic materials reported so far. For examples, the absolute value of $E_{MAE}$ of DyOCl is 58 times bigger than that of Cr$_2$Ge$_2$Te$_6$ (100~$\mu$eV/Cr)\cite{T10} and 8 times bigger than that of CrI$_3$ (655~$\mu$eV/Cr)\cite{T9}. For the $E_{MAE}$ calculations of DyOCl, the initial easy- and hard-axis are set to be $a$- and $c$-axis respectively. So the negative value of $E_{MAE}$ confirms the $a$-axis moment allignment of DyOCl. The above result is in good agreement with the highly anisotropic magnetic properties of DyOCl measured experimentally. It would also be interesting for future theoretical calculations on the magnetic anisotropy energy between $a$- and $[110]$-axis, which should also be a large value according our experimental observations.

The strong uniaxial anisotropy and large $E_{MAE}$ of DyOCl is consistent with an Ising-like single-ion anisotropy of Dy$^{3+}$, which is likely due to the spin-orbit coupling and crystal field effect as in other Dy-based magnetic materials\cite{DyScO3}. Although the moment of Dy lies in the easy $ab$-plane, the strong in-plane anisotropy and magnetic dipolar interactions makes DyOCl a very promising candidate to keep long-range magnetic order down to few layers, according to previous theoretical models\cite{SWT1, SWT2}. So far, most 2D magnetic materials have either out-of-plane anisotropy such as FePS$_{3}$, CrI$_{3}$, VI$_{3}$, Cr$_{2}$Ge$_{2}$Te$_{6}$ and Fe$_{3}$GeTe$_{2}$, or in-plane magnetization but negligible in-plane anisotropy such as CrCl$_3$\cite{CrCl3}. A previous rare example which exhibits substantial in-plane anisotropy is the spin-orbit vdW magnet $\alpha$-RuCl$_3$, which is ascribed to the existence of the off-diagonal $\Gamma$ interaction\cite{RuCl3}. So the strong in-plane $a$-axis anisotropy makes DyOCl another unique example in this research field. In addition, the multi-stage spin-flip transitions and the anisotropic magnetic properties below $T^*$ also create opportunities for exploring novel magnetic phenomena in the 2D limit.

\subsection{Band structure and exfoliation of bulk crystals}

\begin{figure}
	\includegraphics[width=7cm]{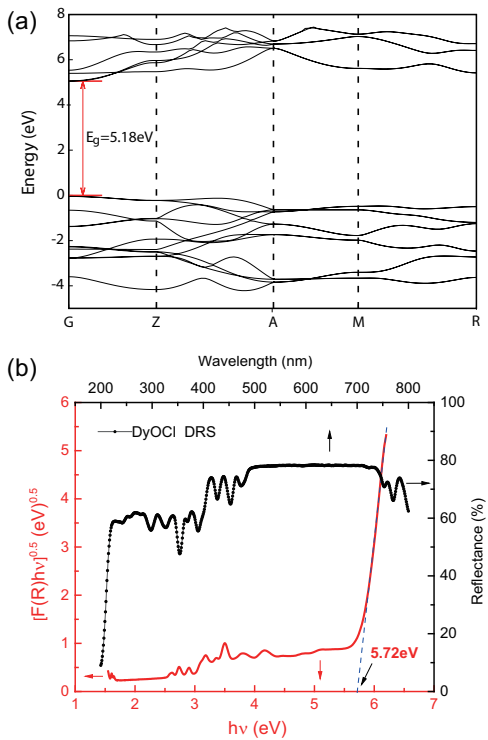}
	\caption {(a) The calculated band structure of DyOCl. (b) The diffusion reflectance spectroscopy (DRS) of DyOCl powder samples (top and right coordinates) and the plot of $[F(R) h\nu ]^{0.5}$ versus photoenergy $h\nu$ (bottom and left coordinates).} \label{fig4}
\end{figure}

The nature of band gap is an important factor for 2D
materials. The band structure of DyOCl obtained by DFT calculations is shown in Fig. 4(a). The result reveals the insulating nature of DyOCl and the calculated direct band gap from $G$ point in valence band maximum to $G$ point in conduction band minimum of DyOCl is $5.18~eV$. In order to verify the calculation results, the diffusion reflectance spectroscopy (DRS) of DyOCl powder samples was measured to determine the band gap. The data is presented in Fig. 4(b) and the absorption band at around $210~nm$ is detected. The plot of $[F(R) h\nu ]^{0.5}$ versus photoenergy $h\nu$ is also shown in the figure. $F(R)$ is the Kubella-Munk function, defined as $F(R)=(1-R)^2/2R$\cite{DRS1}. $R$ is the experimentally observed reflectance. Using the methods proposed in previous publications\cite{DRS1,DRS2}, the optical band gap of DyOCl is calculated to be $5.72~eV$. This experimental value is in good agreement with the DFT calculated value, the difference between them is only $10\%$.  

To demonstrate that DyOCl can be thinned down to few layers because of the existence of vdW gap between adjacent Cl-layers, we performed micro-mechanical exfoliation of DyOCl crystals using Scotch tape. These exfoliated layers on tape were pressed and transferred onto $300~nm$ $SiO_2/Si$ substrates. From the atomic force microscopy image and height profile measurements, thin few-layer flakes of DyOCl with thickness from $4.5~nm$ to $27~nm$ were obtained (Fig. 5). The minimum thickness is $4.5~nm$, which corresponds to about seven layers. The successful exfoliation of DyOCl down to few layers is vital towards future studies of magnetism in 2D limit. Furthermore, this would also allow the incorporating of magnetic ordered DyOCl into various vdW heterostructures and exploring potential use in novel magneto-electronic devices.

\section{Conclusions}

\begin{figure}
	\includegraphics[width=7cm]{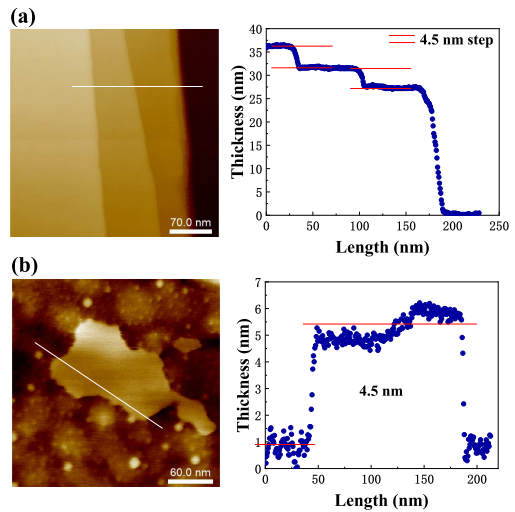}
	\caption {(a) Atomic force microscopy image and height profile step of DyOCl nano-flake mechanically-exfoliated onto a 300 nm SiO$_{2}$/Si substrate. (b) A continuous thin flake of DyOCl with thickness of $4.5~nm$.} \label{fig4}
\end{figure}

In summary, we find DyOCl is a rare-earth based 2D vdW magnetic material with intriguing anisotropic magnetic properties. DyOCl forms an $A$-type AFM structure with the magnetic moment of $10.1~\mu_{B}/Dy$ lying along $a$-axis at $3.5~K$. With magnetic field $H\textgreater 20~kOe$ applied along the easy axis, it can be polarized to ferromagnetic state through spin-flip transition. DFT calculations reveal an extremely large magnetic anisotropy energy of DyOCl ($E_{MAE}$=$ -5850~\mu eV/Dy$), consistent with the strong Ising-like uniaxial magnetic anisotropy observed experimentally. Using mechanical exfoliation, nanoflakes of DyOCl with thickness down to seven layers can be obtained. These properties make DyOCl a promising candidate to realize $f$-electron magnetism in 2D limit. Comparing with the extensively studied transitional metal based vdW magnets, DyOCl provides a new material playground for studying 2D magnetism and its proximate coupling to other 2D materials, exploring novel magneto-electronic devices.

\section*{Acknowledgement}
The authors would like to thank Prof. Yaomin Dai for the help on spectrum data analysis. This work was supported by the National Natural Science Foundation of China  (No. 12074426, No. 11227906 and No. 12004426), the Fundamental Research Funds for the Central Universities, the Research Funds of Renmin University of China (No. 20XNLG19 and No. 21XNLG20), the National Key Research \& Development Projects (No. 2020YFA0406000 and No. 2020YFA0406003).

\bibliography{DyOCl}{}
\end{document}